\documentclass[psfig,12pt]{article}

\usepackage{graphicx}
\usepackage{epstopdf}

\usepackage{color}

\newcommand{\Black}{\color [rgb]{0,0,0}}

\newcommand{\Brown}{\color [rgb]{0.4,0.1,0.1}}

\def\TL{\hfil$\displaystyle{##}$}
\def\TR{$\displaystyle{{}##}$\hfil}
\def\TC{\hfil$\displaystyle{##}$\hfil}
\def\TT{\hbox{##}}
\def\seqalign#1#2{\vcenter{\openup1\jot
  \halign{\strut #1\cr #2 \cr}}}

\def\fixit#1{}

\def\mop#1{\mathop{\rm #1}\nolimits}

\def\Im{\mop{Im}}

\def\overleftrightarrow#1{\vbox{\ialign{##\crcr
     $\leftrightarrow$\crcr\noalign{\kern-0pt\nointerlineskip}
     $\hfil\displaystyle{#1}\hfil$\crcr}}}

\def\lsim{\mathrel{\mathstrut\smash{\ooalign{\raise2.5pt\hbox{$<$}\cr\lower2.5pt\hbox{$\sim$}}}}}
\def\gsim{\mathrel{\mathstrut\smash{\ooalign{\raise2.5pt\hbox{$>$}\cr\lower2.5pt\hbox{$\sim$}}}}}

\def\sqr#1#2{{\vcenter{\vbox{\hrule height.#2pt
         \hbox{\vrule width.#2pt height#1pt \kern#1pt
            \vrule width.#2pt}
         \hrule height.#2pt}}}}
\def\square{\mathop{\mathchoice\sqr56\sqr56\sqr{3.75}4\sqr34\,}\nolimits}

\def\href#1#2{#2}

\def\lbldef#1#2{\expandafter\gdef\csname #1\endcsname {#2}}
\def\eqn#1#2{\lbldef{#1}{(\ref{#1})}%
\begin{equation} \eqalign{#2} \label{#1} \end{equation}}
\def\eqalign#1{\vcenter{\openup1\jot
    \halign{\strut\span\TL & \span\TR\cr #1 \cr
   }}}

\begin{document}
\pagestyle{plain}
\setcounter{page}{1}
\begin{titlepage}

\begin{flushright}
PUPT-2111 \\
UCB-PTH-04/04 \\
LBNL-54593 \\
hep-th/0402156
\end{flushright}
\vfil

\begin{center}
{\huge String creation in cosmologies\\[10pt] with a varying dilaton}
\end{center}

\vfil
\begin{center}
{\large Joshua J. Friess,$^1$ Steven S. Gubser,$^1$ and Indrajit Mitra$^{2,3}$}
\end{center}

$$\seqalign{\span\TL & \span\TT}{
{}^1 & Joseph Henry Laboratories, Princeton University, Princeton, NJ 08544  \cr
{}^2 & Berkeley Center for Theoretical Physics and Department of Physics \cr\noalign{\vskip-1.5\jot}
& University of California, Berkeley, CA 94720-7300 \cr
{}^3 & Theoretical Physics Group, Lawrence Berkeley National Laboratory \cr\noalign{\vskip-1.5\jot}
& Berkeley, CA 94720-8162
}$$
\vfil

\begin{center}
{\large Abstract}
\end{center}

\noindent
FRW solutions of the string theory low-energy effective actions are 
described, yielding a dilaton which first decreases and then increases.  We study string creation in these backgrounds and find an exponential divergence due to an initial space-like singularity.  We conjecture that this singularity may be removed by the effects of back-reaction, leading to a solution which at early times is de Sitter space.

\vfil
\begin{flushleft}
February, 2004
\end{flushleft}
\end{titlepage}
\newpage

\section{Introduction}
\label{INTRODUCTION}

In \cite{clw}, cosmological solutions to the equations of motion derived  
from the NS sector of the string theory effective action were described.  
These solutions are without exception singular at some point in their 
evolution, as they must be, since the starting point actions satisfy the 
hypotheses of the classic singularity theorems of Hawking and Penrose.  
The dilaton in these solutions first decreases and then increases.  In the 
dual language of Horava-Witten theory (if our starting point was the 
perturbative heterotic string), these solutions describe branes which 
start out moving toward one another, reach some minimum distance, and then 
move away again.  This bounce-like behavior is made possible by the 
time-dependence of an axion field, which combines with the dilaton to 
give a single complex scalar.

Such solutions are interesting for several reasons:
 \begin{enumerate}
  \item They are time-dependent backgrounds of string theory, involving only NS fields.  As such, they may admit a CFT description which could address the issue of singularities in a new way.
  \item Such backgrounds offer the possibility of studying string creation.  
  \item The bounce is reminiscent of ekpyrosis \cite{ekpyrosis1,ekpyrosis2}, only there is no singular event like the collision of the branes, and the four-dimensional Einstein frame metric does not proceed from contraction to expansion.
 \end{enumerate} 
In section~\ref{SOLUTIONS}, we will describe a generalization of the results found in 
\cite{clw}, and in subsequent sections we will consider string creation at 
the level of effective field theory, as in \cite{lmCreate,gCreate}.  In a 
particular case, discussed in section~\ref{WHITTAKER} below, the relevant 
wave equation can be solved analytically in terms of Whittaker functions, 
up to terms which do not affect the high mass asymptotics.  In this case, 
there is a divergent number density of strings produced via the cosmic 
expansion, indicating that their back-reaction effects are important.  
Using less 
precise methods, we argue in section~\ref{OTHERK} that this feature persists for other cases, and that the difficulty lies in giving a proper treatment of the initial singularity.

In section~\ref{DISCUSSION}, we attempt to include back-reaction from excited string modes in the equations.  Motivated by the need to keep string creation finite, we suggest that at early times, there may be a de Sitter solution with curvatures on the order of the string scale.  Equations following from the two-derivative action are of limited utility in such a case, but following some clues provided by such equations, we arrive at a proposal for an early time solution which seems to provide a resolution of the early time singularities that plague the geometries explored in earlier sections.

There is a considerable literature on cosmological solutions of string theory.  Rather than present a thorough review of it, we refer the interested reader to \cite{clwReview}.

\section{Solutions with an evolving dilaton}
\label{SOLUTIONS}

The most efficient approach to obtaining the solutions we are interested 
in is to start in eleven dimensions with the action
 \eqn{ElevenDimAction}{
  S_{11} = {1 \over 2\kappa_{11}^2} \int d^{11} x \, \sqrt{g}
   \left[ R - {1 \over 2} G_4^2 \right] + \ldots \,,
 }
where $G_4$ is the anti-symmetric four-form of M-theory and the omitted terms include both the Chern-Simons term for $G_4$ (which will not enter into our considerations) and terms for the gravitinos.  The ansatz to be considered is
 \eqn{ElevenAnsatz}{
  ds^2 &= -e^{2A} d\eta^2 + e^{2B} d\vec{x}^2 + e^{2C} dy_i^2 + e^{2D} dz^2
    \cr
  G_4 &= h dx^1 \wedge dx^2 \wedge dx^3 \wedge dz \,,
 }
where $A$, $B$, $C$, and $D$ are functions only of the time coordinate $\eta$, and $h$ is a constant.  The three coordinates $\vec{x}$ parametrize the three large spatial dimensions; the six coordinates $y_i$ parametrize a compact Ricci-flat manifold; and $z$ is the eleventh dimension, most simply taken to be a circle.\footnote{Given a solution of the form we have described, we could orbifold by ${\bf Z}_2$ to get an approximate solution to Horava-Witten theory.  It is only approximate for most choices of the internal manifold parametrized by the $y_i$ because usually one has some $G_4$ on the internal six-manifold.  Then the Chern-Simons term for $G_4$ would matter, and the functions in \ElevenAnsatz\ would depend at the least on $\eta$ and $z$.}  Both the Bianchi identity for $G_4$ and its equation of motion are trivially satisfied.  The solutions of the equations of motion may be conveniently expressed as follows:
 \eqn{FRWsolns}{
  A &= 3B + D  \cr
  B &= B_0 + {1 \over 3} \log\sec h\eta + 
   {q \over 2} \log {1 + \sin h\eta \over \cos h\eta}  \cr
  C &= C_0 + {1 \over 6} \log\cos h\eta  \cr
  D &= D_0 + {1 - 3 q^2 \over 6 q} \log {1 + \sin h\eta \over \cos h\eta} - 
   {1 \over 3} \log \cos h\eta \,,
 }
where $B_0$, $C_0$, $D_0$, and $q \neq 0$ are real constants.  The variable $\eta$ 
runs from $-\pi/2h$ to $\pi/2h$, and there is a curvature singularity at 
one of these endpoints.  Choosing $q > 0$ puts that singularity at $\eta = 
-\pi/2h$; changing the sign of $q$ corresponds to time reversing the 
entire solution.

Reducing the solution \FRWsolns\ on the circle parametrized by $z$ results in a solution to the type IIA low-energy effective action that depends only on NS fields:
 \eqn{ReductionToIIA}{\seqalign{\span\TC}{
  ds_{11}^2 = e^{-{2 \over 3} \Phi} ds_{str}^2 +  
   e^{{4 \over 3} \Phi} dz^2  \qquad
  G_4 = H_3 \wedge dz  \cr
  ds_{str}^2 = -e^{2\alpha} d\eta^2 + e^{2\beta} d\vec{x}^2 + 
    e^{2\gamma} dy_i^2  \cr
  \alpha = A + {D \over 2} \quad \beta = B + {D \over 2} \quad 
   \gamma = C + {D \over 2} \quad
  \Phi = {3D \over 2} \quad H_3 = h dx^1 \wedge dx^2 \wedge dx^3
 }}
The reduction to four dimensions is trivial in string frame, except that one must define a new dilaton in four dimensions:
 \eqn{FourDdilaton}{
  \phi = 2\Phi - {1 \over 2} \log \det G_{ij} = -6C = 
   -6 C_0 - \log \cos h\eta \,,
 }
where $\det G_{ij} = e^{12 \gamma}$ is the determinant of the string frame metric in the six compact directions parametrized by $y_i$.  The four-dimensional Einstein metric is
 \eqn{FourDEinstein}{\seqalign{\span\TC}{
  ds_{4E}^2 = e^{-\phi} ds_{4,str}^2 = 
   -e^{2A'} d\eta^2 + e^{2B'} d\vec{x}^2 \cr
  A' = A + {D \over 2} + 3C \qquad
  B' = B + {D \over 2} + 3C \,.
 }}

It will also prove useful to write the solution in terms of a different 
time variable, $\tau$, defined by 
 \eqn{dTauDef}{\seqalign{\span\TC}{
  d\tau = e^{\alpha-\beta} d\eta \qquad 
  h\tau = K \left( {1+\sin h\eta \over \cos h\eta} \right)^{1/K}  \cr
  K \equiv {6q \over 1+3q^2} \,,
 }}
where we have set $B_0 = C_0 = D_0 = 0$.  This can be accomplished by
rescaling coordinates in the eleven-dimensional solution. In terms of this new time variable, the four
dimensional Einstein frame metric \FourDEinstein\ takes on a
particularly simple form:
\eqn{SimpleEinstein}{
  ds_{4E}^2 = \left( { h \tau \over K} \right) \left[ -d \tau^2 +
    d\vec{x}^2 \right] \,.
}
This is an expected result, because in four-dimensional Einstein frame, the dynamics is described by gravity coupled to a non-linear sigma model with zero potential.  The sigma model supplies stress energy with $w=1$, and \SimpleEinstein\ is the $k=0$ FRW solution with such stress energy.

The ten-dimensional string metric may be written as
 \eqn{StringMetricAgain}{
  ds_{str}^2 &= a^2 \left( -d\tau^2 + d\vec{x}^2 \right) + 
   b^2 dy_i^2  \cr
  a^2 &= {1 \over 2} \left[ \left( {h\tau \over K} \right)^K + 
   \left( {h\tau \over K} \right)^{-K} \right] {h\tau \over K}  \cr
  b^2 &= \left( h\tau \over K \right)^{\pm \frac{1}{K} \sqrt{1-K^2/3}} \,,
 }
where the sign corresponds to the one in the relation $q = K \mp
\sqrt{K^2-1/3}$. One of the solutions discussed in 
\cite{clw} is the case $q = 1/K = 1/\sqrt{3}$ of the solutions presented here. The fractional power-law dependence of the scale factor $b$ tells us
that time $ 0 \leq \tau < \infty$.

\section{Wave equations}
\label{WAVE}

To study string propagation in the backgrounds constructed in section~\ref{SOLUTIONS}, the ideal approach would be to use the freedom to shift the zero point of the dilaton to make the string coupling very small except at the earliest and latest of times; to make some reasonable assumptions about in and out states at these times; and to employ a worldsheet description to compute the evolution while the string coupling is small.  We will accomplish less than that in this paper: our starting point will be the scalar wave equation
 \eqn{ScalarEOM}{
  \left( -\square + m^2 + \xi R \right) \phi = 0 \,.
 }
Individual components of higher spin fields satisfy such equations, with values of $\xi$ and $m$ depending on the properties of 
the field in question.  The on-shell condition $L_0 |\Psi\rangle = 0$ for a string state will lead approximately to an equation 
of the form \ScalarEOM.  For an excited string, the $m^2$ term dominates over $\xi R$ unless $R$ becomes of order the string 
scale, at which point a purely geometric description breaks down.  In the singular solutions constructed in 
section~\ref{SOLUTIONS}, $R$ is unbounded at early times, so this break-down occurs.  We will nevertheless investigate 
solutions of \ScalarEOM\ for special values of $\xi$ and the parameter $q$ introduced in section~\ref{SOLUTIONS}.  


Using a separation of variables ansatz,
 \eqn{SeparateVariables}{
  \phi = {1 \over ab^3} e^{i \vec{k} \cdot \vec{x} + iv_j y^j}
   \chi(\tau) \,,
 }
one derives from \ScalarEOM\ the equations
 \eqn{TwoForms}{\seqalign{\span\TC}{
  \left[ \partial_\tau^2 + \vec{k}^2 + m^2 a^2 +
   {a^2 \over b^2} v_j^2 - {1 \over ab^3} (\partial_\tau^2 ab^3) + 
   \xi R a^2 \right]
    \chi = 0  \cr
  \left[ \partial_\tau^2 + \vec{k}^2 + m^2 a^2 +
   {a^2 \over b^2} v_j^2 + (6\xi-1) {\ddot{a} \over a} + 
   3 (4\xi-1) {\ddot{b} \over b} + 6 (5\xi-1) {\dot{b}^2 \over b^2} + 
   6 (4\xi-1) {\dot{a} \dot{b} \over ab} \right] \chi = 0 \,,
 }}
where the second line follows from the first plus an expression for the string-frame Ricci scalar $R$ in ten 
dimensions.  We have assumed that the $y_i$ directions form a torus, but the analysis would be no different for 
a more general six-manifold: $-v_j^2$ would still be an eigenvalue of the laplacian on the six-manifold.

\section{An analytic approximation for $K=1$}
\label{WHITTAKER}

Equation \TwoForms\ is complicated for general $K$.  For the choice $K=1$, however, there is an approximate solution which improves as the mass is increased, and using this solution one can calculate the expected string production in a fairly straightforward manner.  The fully explicit form of \TwoForms\ for $K=1$ (with $v_j = 0$) is
 \eqn{LastForm}{\seqalign{\span\TC}{
   \left[ \partial_\tau^2 + \Omega^2(\tau) \right] \chi = 0
    \qquad\hbox{where}  \cr
   \Omega^2(\tau) = 
  k^2 + \frac{m^2}{2}(1 + h^2\tau^2) + \xi_1 \frac{1}{\tau^2} +
  \xi_2 \frac{h^2}{1 + h^2 \tau^2} + \xi_3 \frac{h^2}{(1 + h^2\tau^2)^2} \,,
 }}
and 
 \eqn{xis}{
  \xi_1 &= (7-2\sqrt{6})\xi + \sqrt{\frac{3}{2}} - \frac{3}{2} \cr
  \xi_2 &= \sqrt{6}(4\xi -1) \cr
  \xi_3 &= 6\xi - 1  \,.
 }
Note that the last two terms in $\Omega^2(\tau)$ are each bounded between 0 
(at late times) and $h^2$ (at early times).  At very early times, the $1/\tau^2$ term dominates both of the last two terms.  At late times, the $m^2h^2\tau^2$ term dominates.  During intermediate times, when $\tau \sim 1/h$, all of these terms are of nearly the same order.  However, for sufficiently large masses (that is for $m^2 \gg h^2$), the $m^2$ term will dominate all of them.  Hence we can drop the last two terms for large mass strings, which gives us
 \eqn{ApproxWave}{
  \left[\partial_\tau^2 + k^2 + \frac{m^2}{2}(1 + h^2\tau^2) + \xi_1
  \frac{1}{\tau^2}\right]\chi \approx 0  \,.
 }
The solutions to \ApproxWave\ are exact, and are given in terms of the Whittaker
functions, $M_{\lambda,\mu}(z)$ and $W_{\lambda,\mu}(z)$:
 \eqn{solutions}{
  \chi_1(\tau) & = \frac{1}{\tau^{1/2}}
    M_{\lambda,\mu}\left(\frac{imh\tau^2}{\sqrt{2}}\right) \cr
  \chi_2(\tau) & = \frac{1}{\tau^{1/2}}
    W_{\lambda,\mu}\left(\frac{imh\tau^2}{\sqrt{2}}\right)  \,,
 }
where
 \eqn{lambdamu}{
  \lambda & = \frac{-i (k^2 + m^2/2)}{2\sqrt{2} mh} \qquad
  \mu = \frac{\sqrt{1-4\xi_1}}{4} \,.
 }
Both $M$ and $W$ can be written in terms of confluent hypergeometric functions
 \eqn{MWDef}{
  M_{\lambda,\mu}(z) & = z^{1/2 + \mu} e^{-z/2} 
   \Phi(1/2 - \lambda + \mu, 2\mu + 1; z) \cr
  W_{\lambda,\mu}(z) &= z^{1/2 + \mu} e^{-z/2} 
   \Psi(1/2 - \lambda + \mu,2\mu + 1; z) \,,
 }
where $\Phi$ and $\Psi$ are confluent hypergeometric functions of the first and second kind, respectively. It is important to note that the Whittaker functions $M$ and $W$ are linearly independent, but they are not orthogonal.  

The computation of particle production proceeds by first selecting natural asymptotic in and out vacua. Our choice of
vacua will be dictated by the criterion that we have
oscillatory modes with well-defined positive frequency. We shall use a
generalization of the flat-space definition of positive frequency,
which requires that these modes satisfy the relation 
\eqn{Genfreq}{
 -\Im\left\{ \partial_\tau \log\phi \right\} > 0 \,.
}
In time-dependent backgrounds, the in and out vacua are different:
this means that if the universe begins in the
in-vacuum state, then at late times, the universe will be in a state which 
is excited with
respect to the out-vacuum; in other words, particles are produced as a 
result of the expansion. We would like 
to find the linear combinations of the $\chi_i$'s that serve as the positive
frequency solutions at asymptotically early and late times. Recall
that in our case, $z = 
imh\tau^2$; early times correspond to $\tau \rightarrow 0$, and late times correspond to $\tau
\rightarrow \infty$. 

The leading term in the small $z$ expansion of $\Phi$ is constant;
hence for $z \to 0$, we have $M_{\lambda,\mu}(z) \sim z^{1/2 + \mu}
e^{-z/2}$.  The requirement of oscillatory behavior at 
very early times (corresponding to $\tau \sim 0$) forces $\mu$ to be
imaginary: from \lambdamu\ we see that this requires $\xi_1 >
1/4$. The oscillation of $M_{\lambda, \mu}$ is then dominated by the 
$z^\mu$ factor. Therefore, $\chi_1^*(\tau)$ is the time-dependent piece of 
the early-time positive 
frequency mode, and hence can be used to define a good in-vacuum upon 
quantization:

\eqn{instate}{
 \phi_{in} = {N_{in} \over ab^3} e^{i \vec{k} \cdot \vec{x} + iv_j y^j}
   \chi_1^*(\tau) \,,
}
where we have included an undetermined normalization factor $N_{in}$ for this 
mode.

To choose the out-vacuum, we examine the large $z$ expansion of the
function $W_{\lambda,\mu}(z)$. The leading term in the large
$z$ expansion of $\Psi$ is $\Psi \sim z^{-\alpha}$, for $\alpha = 1/2 - \lambda + \mu$.  Hence for
large $z$, functions proportional to $W_{\lambda,\mu}(z)$ will have definite frequency,
dominated by the $e^{-z}$ factor. So we take $\chi_2(\tau)$ as the positive frequency mode that defines our
out-vacuum:
\eqn{outstate}{ 
 \phi_{out} = {N_{out} \over ab^3} e^{i \vec{k} \cdot \vec{x} + iv_j y^j}
   \chi_2 (\tau) \,.
}

The Boglubov coefficients, which determine how much the in-vacuum differs from the out-vacuum, can be determined by expressing
\eqn{Bogoliubov}{
 \phi_{in} = \alpha \phi_{out} + \beta \phi_{out}^* \,.
}
To determine $\alpha$ and $\beta$, we need to know how the functions
$M$ and $W$ are linearly related. For $- \pi /2 < \arg(z) < 3 \pi /2$ the
following relation holds (see for instance
\cite{GR} and \cite{lebedev}):
\eqn{MinW}{
M_{\lambda,\mu}(z) = \frac{\Gamma(1+2\mu)}{\Gamma(1/2 + \mu - \lambda)} e^{-i \pi \lambda}
W_{-\lambda, \mu}(-z) + \frac{\Gamma(1+2\mu)}{\Gamma(1/2 + \mu + \lambda)} e^{-i \pi(\lambda - \mu
- 1/2)} W_{\lambda, \mu}(z) \,.
}
To be able to use this relation to compute $\alpha$ and $\beta$, one
has to use the complex conjugate of \Bogoliubov\ (this ensures that
$\arg (z) = \pi /2$) :
\eqn{LinDecomp}{\eqalign{
 \phi_{in}^* &= N_{in}^* \left[ \frac{\Gamma(1+2\mu)}{\Gamma(1/2 + \mu - \lambda)} e^{-i \pi \lambda}
W_{-\lambda, \mu}({{-i mh \tau^2} \over {\sqrt 2}}) + \frac{\Gamma(1+2\mu)}{\Gamma(1/2 + \mu + \lambda)} e^{-i \pi(\lambda - \mu
- 1/2)} W_{\lambda, \mu}({i mh \tau^2 \over {\sqrt 2}}) \right] \cr 
 &= {N_{in}^* \over N_{out}^*} \frac{\Gamma(1+2\mu)}{\Gamma(1/2 + \mu
   - \lambda)} e^{-i \pi \lambda} \phi_{out}^* + {N_{in}^* \over
   N_{out}} \frac{\Gamma(1+2\mu)}{\Gamma(1/2 + \mu + \lambda)} e^{-i \pi(\lambda - \mu
- 1/2)} \phi_{out} \,,
}}
where in the last line we have used the general property $W_{\lambda, \mu}(z) =
W_{\lambda,-\mu}(z)$ and the fact that for us $\lambda$, $\mu$, and
$z$ are all pure imaginary. This allows us to read off $\alpha$ and $\beta$, up to
some relative normalization of the wave functions: 
\begin{eqnarray}
\alpha^* & = & {N_{in}^* \over N_{out}^*} \frac{\Gamma(1+2\mu)}{\Gamma(1/2 - \lambda + \mu)} e^{-i\pi\lambda} \\
\beta^* & = &  {N_{in}^* \over N_{out}} \frac{\Gamma(1+2\mu)}{\Gamma(1/2 + \lambda + \mu)}
e^{-i\pi(\lambda - \mu - 1/2)} \,.
\end{eqnarray}
Using the unitarity constraint $|\alpha|^2 - |\beta|^2 = 1 $ to fix
the normalization we find our desired result:\footnote{We essentially choose $|\mu|$ to be any positive real
number, based on our choice of $\xi_1 > 1/4$.}

\eqn{FinalBeta}{
|\beta|^2 &= \frac{1}{2 \sinh(2\pi|\mu|)} \left[ e^{-2\pi|\lambda|} + e^{-2\pi|\mu|} \right] \cr
|\lambda| &= \frac{(k^2 + m^2/2)}{2\sqrt{2} mh} \,.
}

The main qualitative point to note is that the particle production probability $|\beta|^2$ approaches some asymptotically constant, non-zero value as $m\rightarrow 0$ and
$m\rightarrow \infty$.  But the number of 
string modes increases exponentially 
with mass, and each mode experiences pair-creation independently, provided 
that the string coupling is weak.  Thus we conclude that the $K=1$ case 
suffers from the problem of divergent string production, at least 
before back-reaction of the massive modes is properly taken into account.

\section{Other values of $K$}
\label{OTHERK}

The range of possible values for $K$ is $0 < K \leq \sqrt{3}$.  The 
original $d=4$ solution of \cite{clw} is the special case $K=\sqrt{3}$.  
In 
this section we will argue that, at least for a wide range of the 
parameter $\xi$, the particle occupation number $|\beta|^2$ is finite at 
large mass. Provided the Bogolubov coefficient $\beta$ is small, there is a useful approximate formula,
 \eqn{MasterBeta}{
  \beta \approx \int d\tau \, {\dot\omega \over 2\omega}
   \exp\left( -2i \int^\tau du \, \omega(u) \right) \,,
 }
where the integral is over the whole range of $\tau$---in our case, $0$ to $\infty$.  The result \MasterBeta\ can be derived by expressing a general solution to the separated wave equation \LastForm\ in terms of zeroeth order WKB solutions times slowly varying coefficients.  The formula \MasterBeta\ is the starting point for the method of steepest descent used in \cite{lmCreate,gCreate}.  This method relies however on the range of $\tau$ being $-\infty$ to $\infty$, and it requires some regularity properties of the integrand which amount to having well-defined adiabatic in and out vacua.  Neither of these conditions pertains to the current problem.  However, as we saw in section~\ref{WHITTAKER}, there can nevertheless be a natural choice of in and out vacua, and finite results for $\beta$ may be obtained.  To get a more general view on the problem, let us choose some $\tau_0$ and define
 \eqn{DefineX}{
  x(\tau) = \int_{\tau_0}^\tau du \, \omega(u) \,.
 }
The variable $x$ is a monotonically increasing function of $\tau$, and we 
will assume that its range is $-\infty$ to $\infty$.  This assumption tends to lead to naturally defined in and out vacua because there is a solution to the separated wave equation \TwoForms\ whose phase is approximately $e^{-i x(\tau)}$ for early times, and another with the same property at late times.  These solutions are naturally identified as positive frequency modes at early or late times.

One may rewrite \MasterBeta\ as 
 \eqn{MasterBetaX}{\seqalign{\span\TC}{
  \beta \approx \int_{-\infty}^\infty dx \, e^{-2ix} \Xi(x) 
   \qquad\hbox{where}\quad
  \Xi = {d\omega/d\tau \over 2\omega^2} \,.
 }}
The smallness $\Xi$ is a standard measure of the reliability of the WKB approximation.  Evidently, the integral \MasterBetaX\ converges absolutely provided $\Xi(x)$ is itself integrable.  If $\Xi(x)$ is not integrable due to behavior near $\pm\infty$, we must regulate the integral.  Suppose there are early time and late time expansions of the form
 \eqn{LateTimeExpand}{\seqalign{\span\TL & \span\TR &\qquad\span\TT &\quad \span\TR}{
  \Xi &= \sum_j c^{(i)}_j (-x)^{a_j} & for & x \to -\infty  \cr
  \Xi &= \sum_j c^{(f)}_j x^{b_j} & for & x \to \infty \,.
 }}
Then, given some $x_i$ and $x_f$, we may evaluate \MasterBetaX\ as
 \eqn{SplitBeta}{
  \beta &\approx \int_{-\infty}^{x_i} dx \, e^{-2ix}
    \sum_j c^{(i)}_j (-x)^{a_j} + 
   \int_{x_i}^{x_f} dx \, e^{-2ix} \Xi(x) + 
   \int_{x_f}^\infty dx \, e^{-2ix} \sum_j c^{(f)}_j x^{b_j}  \cr
   &= \int_{x_i}^{x_f} dx \, e^{-2ix} \Xi(x) +
    \sum_j {c^{(i)}_j \over (-2i)^{1+a_j}} \Gamma(1+a_j,2i x_i) +
    \sum_j {c^{(f)}_j \over (2i)^{1+b_j}} \Gamma(1+b_j,2i x_f) \,,
 }
where $\Gamma(a,z)$ is the incomplete gamma function.  If quantitatively precise results are desired, the integral from $x_i$ to $x_f$ can be done numerically and the early and late time series truncated to some finite order.

Let us reexamine the separated wave equation \TwoForms.  If $m$ is large, then the equation is approximately
 \eqn{TwoFormsAgain}{\seqalign{\span\TC}{
  \left[ \partial_\tau^2 + m^2 a^2 + {a^2 \over b^2} v_j^2 + 
   {\xi_1 \over \tau^2} \right] \chi = 0  \cr
  \xi_1 = \left({3 \over 2} K^2 + {21 \over {2 K^2}} - 5 - 
    2 {\sqrt{9 - 3K^2}} \right) \xi + 
    \left( 1 - {9 \over {4 K^2}} - {K^2 \over 4} + {1 \over 2}
    {\sqrt{9 - 3K^2}} \right)
 }}
where we have discarded terms which are smaller than some of the ones we have kept.  If we set $v_j=0$, then we may simplify even further (again discarding terms that are small, for large $m$, compared to some that we keep):
 \eqn{SimplestForm}{
  \left[ \partial_\tau^2 + 
   {m^2 \over 2} \left( \left({h\tau \over K} \right)^{1+K} +\left({h\tau \over K} \right)^{1-K} \right) + 
   {\xi_1 \over \tau^2} \right] \chi = 0 \,.
 }
The second and fourth terms in square brackets are equal when
 \eqn{TauZeroDef}{
  \tau = \tau_0 \equiv 
   \left( {2 K^{K+1} \xi_1 \over h^{K+1} m^2} \right)^{1 \over 3+K}
    \,,
 }
where we have chosen some specific $\tau_0$ for convenience in the next 
few equations.

Asymptotic expressions for $\Xi$ at early and late times are easy to obtain:
 \eqn{GenKStep}{
   \Xi \approx \left\{
    \eqalign{
     -{1\over 2\sqrt{\xi_1}} &\qquad\hbox{for $\tau \ll \tau_0$}  \cr
     {{(1 + K) K^{{1+K} \over 2}} \over 
      {2 \sqrt{2} m h^{{1+K} \over 2} \tau^{{3+K} \over 2}}} 
       &\qquad\hbox{for $\tau \gg \tau_0$.}}
      \right.
 }
As a crude approximation, we may take 
 \eqn{CrudeXi}{
  \Xi \approx -{1 \over 2\sqrt{\xi_1}} \theta(x) \,,
 }
where $\theta(x)$ is the usual step function.  Using \CrudeXi\ and \SplitBeta\ leads immediately to
 \eqn{CrudeBeta}{
  \beta \approx {1 \over 4i\sqrt{\xi_1}} \,.
 }
The result \CrudeBeta, coming only from the leading early time
 remainder term in \SplitBeta, agrees qualitatively with the analysis
 in section~\ref{WHITTAKER}: for large $m$, $|\beta|^2$ tends to a
 constant, and that constant is small when $\xi_1$ is large.  The
 disagreement on the value of the constant is likely due to the fact
 that the WKB method is being pushed beyond its realm of validity. Clearly, the result \CrudeBeta\ again results in exponentially
divergent total string production.

\section{Resolving the initial singularity}
\label{DISCUSSION}

The finding that the total string production is infinite due to the presence of an initial spacelike curvature singularity may be unsurprising, though it underscores the subtleties which one must face in discussing string dynamics in the early universe.  In the spirit of \cite{lmCreate}, we would like to inquire whether string creation near the spacelike singularity hints at how this singularity might be removed.  Back-reaction is clearly crucial to this story.

Curiously, when one entirely neglects the gravitational back-reaction of the dilaton and the axion on the four-dimensional Einstein frame metric (which is then taken to be flat Minkowski space) one arrives at a wave equation that gives exponentially small production probability $|\beta|^2$ for large $m$.  The details of this have already appeared in \cite{gTalk}, and we will not repeat them here: instead, let us simply quote the result (using our current notation) that for the case $K=\sqrt{3}$ where the extra dimensions are not changing their size in ten-dimensional string frame, the no-back-reaction approximation gives finite total string production when
 \eqn{DilatonEstimate}{
  e^{\phi_0} \gsim {2\pi\alpha' h^2} \,.
 }
Evidently, the difficult situation we have encountered in this paper, where the production probability $|\beta|^2$ is constant at large $m$, owes to the fact that the early time behavior of the solution including back-reaction from the dilaton and the axion is radically different from the no-back-reaction treatment that led to \DilatonEstimate.  How then, we would like to ask, might the inclusion of back-reaction from strings modify the story?\footnote{The construction of non-singular FRW cosmologies as solutions to the string equations including matter has also been studied recently in \cite{tsujikawa}.  The starting assumptions regarding the stress-energy from strings are somewhat different from ours, and consequently the results also differ.}

In \cite{vt} a formalism was introduced to study this sort of question for homogenous isotropic cosmologies.  Let us consider the ansatz
 \eqn{TenAnsatz}{\eqalign{
  ds_{str}^2 &= -dt^2 + e^{2\beta} d\vec{x}^2 + e^{2\gamma} dy_i^2
    \cr
  H_3 &= h dx^1 \wedge dx^2 \wedge dx^3 \qquad \Phi = \Phi(t) \,,
 }}
where as usual the three directions $x^a$ are non-compact, while the six extra dimensions parameterized by the $y^i$ are compact.  One starts with an action
 \eqn{vtAction}{
  S = \int dt \, \sqrt{-G_{00}} \left[
   e^{-\varphi} \left( -G^{00} 3 \dot\beta^2 - G^{00} 6 \dot\gamma^2 +
    G^{00} \dot\varphi^2 + {h^2 \over 2} e^{-6\beta} \right) + 
    F(\beta,\gamma,T\sqrt{G^{00}}) \right] \,,
 }
where all terms but the last are a straightforward reduction of the ten-dimensional string action, $\alpha'=1$, and we have defined
 \eqn{varphiDef}{
  \varphi = 2\Phi - 3\beta - 6 \gamma \,.
 }
The last term is included to describe the strings, which in \cite{vt} were taken to be at finite temperature $T$.  The equations of motion, in the gauge $G_{00} = -1$, are
 \eqn{vtEOMS}{\eqalign{
  \ddot\beta - \dot\varphi \dot\beta - {1 \over 2} h^2 e^{-6\beta} &= 
   {1 \over 2} e^\varphi P_x  \cr
  \ddot\gamma - \dot\varphi \dot\gamma &= {1 \over 2} e^\varphi P_y  \cr
  \ddot\varphi - {1 \over 2} \dot\varphi^2 - {3 \over 2} \dot\beta^2 -
   3 \dot\gamma^2 + 
   {h^2 \over 4} e^{-6\beta} &= 0  \cr
  -3\dot\beta^2 - 6 \dot\gamma^2 + \dot\varphi^2 - 
   {h^2 \over 2} e^{-6\beta} &= e^\varphi E \,,
 }}
where we have defined
 $$
  E = F - T {\partial F \over \partial T} \qquad
  3P_x = -{\partial F \over \partial \beta} \qquad
  6P_y = -{\partial F \over \partial \gamma}  \,.
 $$
$E$ is the total energy in some fixed coordinate volume, and likewise $P_x$ and $P_y$ are extensive quantities.  There is a conservation equation,
 $$
  \dot{E} + 3 \dot\beta P_x + 6 \dot\gamma P_y = 0 \,,
 $$
which follows as a consistency condition of \vtEOMS.  If $E=P_x=P_y=0$, then one can recover the solution \ReductionToIIA.

Motivated by the Brandenberger-Vafa scenario \cite{bv}, let us assume that $\dot\gamma = P_y = 0$: that is, the extra dimensions are stabilized, perhaps at the self-dual radius.  The remaining independent equations of motion can be expressed as
 \eqn{RemainingEOMS}{\eqalign{
  \ddot\varphi - {1 \over 2} \dot\varphi^2 - {3 \over 2} \dot\beta^2 + 
   {h^2 \over 4} e^{-6\beta} &= 0  \cr
  -3\dot\beta^2 + \dot\varphi^2 - 
   {h^2 \over 2} e^{-6\beta} &= e^\varphi E  \cr
  \dot{E} + 3 \dot\beta P_x = 0  \,. 
 }}
If we knew $w=P_x/E$, then these equations could be solved for $\varphi$, $\beta$, and $E$, given initial conditions.  But the stress tensor of strings (insofar as it exists) is difficult to evaluate in a regime where quantum creation of highly excited strings is important, so we clearly do not know $w$.

We do know, however, that strings can provide negative pressure: if strings are stretched further than the Hubble length, then roughly speaking they hold their shape in co-moving coordinates, expanding as the universe expands, and leading to $w = -1/3$.  Quantum creation of strings might also provide negative pressure, since the quantum-created modes add energy as the universe expands.

Negative pressure suggests that perhaps the non-compact part of the string frame metric should be $dS_4$, or something close to it, at early times.  We are interested in describing a regime where string creation is important, and this suggests that the Hubble constant should be related to the Hagedorn temperature of the string by $H/2\pi \approx T_H$.  If the Hubble constant is significantly smaller, then string creation is insignificant, and it is hard to see how exponential expansion would continue.  If the Hubble constant is larger, then it would seem that string creation must be divergent.

Thus, based only on the qualitative consideration of what sort of geometry would lead to significant but finite $E$ from excited strings, we seem drawn to $dS_4$.  Unfortunately, we run into difficulties when we plug the ansatz $\beta = Ht$ into \RemainingEOMS.  From the first of these equations, one obtains a solution for $\varphi$ whose behavior at early times is $\varphi \sim -{h \over 3\sqrt{2} H} e^{-3\beta}$.  This leads to extremely negative $E$, which seems nonsensical.  The origin of the problem is that the terms proportional to $h^2 e^{-6\beta}$ dominate at early times.  It does not make sense physically for these terms to exceed the string scale: $\alpha'$ corrections would set in, and the starting point action would include more than just $-{1 \over 2} H_3^2$.  It makes more sense to assume that, at early times, the contribution from $-{1 \over 2} H_3^2$ to \RemainingEOMS\ is nearly constant, resulting in
 \eqn{ModifiedEOMS}{\eqalign{
  \ddot\varphi - {1 \over 2} \dot\varphi^2 - 
   {3 \over 2} H^2 + {h_\infty^2 \over 4} &= 0  \cr
  -3H^2 + \dot\varphi^2 - {h_\infty^2 \over 2} &= 
   e^\varphi E  \cr
  \dot{E} + 3H P_x = 0 \,,
 }}
for some constant $h_\infty$ which we cannot determine without better knowledge of the stringy corrections.  The first of these equations can be solved for $\varphi$:
 \eqn{varphiSoln}{
  \varphi = \varphi_0 - 2 \log \cosh t/t_0 \qquad
  \hbox{where $1/t_0 = 
   \sqrt{{1 \over 8} h_\infty^2 - {3 \over 4} H^2}$.}
 }
If $t_0$ is real, the solution is well-behaved for all $t$, but it turns out again that $E$ is extremely negative for early times.  If $t_0$ is imaginary, then $E$ is positive for a certain range of $t$, but $\varphi$ becomes singular in finite time.  We should not take seriously a solution in which $\dot\varphi$ exceeds the string scale: $\alpha'$ corrections inevitably become important.  Assuming that such corrections cause $\dot\varphi$ to saturate to some finite negative value, $\dot\varphi_\infty$, at early times---taking the sign from the singular solution \varphiSoln\ with imaginary $t_0$---one finds from the second equation of \ModifiedEOMS\ that
 \eqn{ModifiedAgain}{
  -3H^2 + \dot\varphi_\infty^2 - {h_\infty^2 \over 2} = 
   e^{\varphi_0 + \dot\varphi_\infty t} E \,.
 }
The first equation in \ModifiedEOMS\ should not be considered on a par with the second, because the first follows from functionally differentiating with respect to $\varphi$, and we have assumed that $\alpha'$ corrections substantially change the action's dependence on $\varphi$.  The second, which follows from time-reparametrization invariance, should be a better guide to the string-scale physics.
Assuming the left hand side of \ModifiedAgain is positive, one obtains
 \eqn{NiceE}{
  E = e^{-\dot\varphi_\infty t} E_0 \,,
 }
where $E_0$ is some positive constant.  Recall that $E$ is the total energy in some fixed coordinate volume: $E e^{-3\beta}$ is proportional to the energy density $\rho$ from excited strings.  $E$ is increasing with time, which means that pressure is indeed negative: the conservation equation in \ModifiedEOMS\ leads directly to 
 \eqn{NiceW}{
  w = P_x/E = \dot\varphi_\infty/3H \,,
 }
which is constant and negative.  From \ModifiedAgain\ one may read off $w < -1/\sqrt{3}$, which means that the co-moving expansion of highly excited strings cannot be the only source of negative pressure.  This is at least consistent with the observation that quantum creation of strings causes the total energy in string modes to increase.  Using \varphiDef\ and \NiceW, one can show $\dot\Phi = {3H \over 2} (w+1)$.

In summary, the ``solution'' described around \ModifiedAgain\ passes all the consistency checks that we can make.  It describes $dS_4$ both in string frame and Einstein frame because the dilaton varies linearly with time.  It is really only a conjecture because we have extrapolated beyond the starting point equations \vtEOMS, assuming that various terms in these equations saturate at finite values of order the string scale.  The freedom to shift the dilaton by an additive constant remains, so there can be a large region of the solution where higher loop corrections can be neglected.

\begin{figure}[t]
  \centerline{\includegraphics[width=3in]{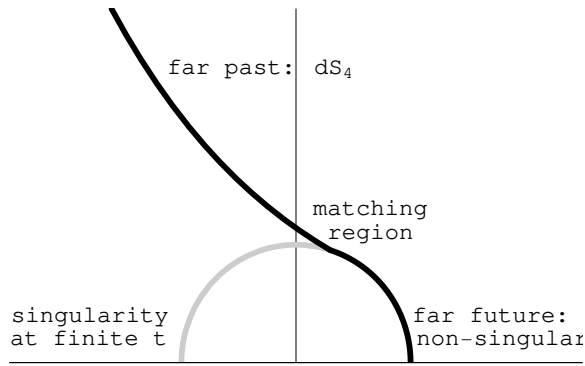}}
  \caption{The conjectured form of a matched solution for $w>-1$ and $K=\sqrt{3}$.  The grey line represents the original solution~\ReductionToIIA\ at early times, leading back to a space-like singularity.}\label{figA}
 \end{figure}

Matching onto the solutions described in section~\ref{SOLUTIONS} at early times is complicated by the fact that we don't know the sign of $\dot\Phi$, and $\dot\beta$ is positive in the $dS_4$ solution but negative for $K>1$ in the solutions of section~\ref{SOLUTIONS}.  If $w>-1$ and $K=\sqrt{3}$, then the matching point in the solution described in \ReductionToIIA\ might be expected to be for $\eta>0$, where $\dot\beta>0$ and $\dot\Phi>0$.  Then in the plane of the complexified dilaton, $\chi + i e^{-\phi}$, where $d\chi = e^{-\phi} * H_3$ in four-dimensional Einstein frame, the total solution would look something like the curve drawn in figure~\ref{figA}.

Clearly it would be desirable to investigate the conjectured de Sitter solution with better input from string worldsheet physics.  This is not an easy task because the solution is largely supported by stress-energy from highly excited strings that are produced by quantum effects as the universe expands.  It is possible that a space-time description is wholly inadequate to describe the physics: curvatures on the order of the string scale are a hallmark of the breakdown of spacetime physics.  However, the solution we have suggested seems so uniquely suited to avoiding the difficulties of exponentially divergent string production that we expect it to be at least a rough guide to the true early time physics.  If so, then of course one would ask if there is a connection to be made with inflation.  This would require the string coupling to be extremely weak at early times, so that the string scale is a factor of $10^{-5}$ or so below the Planck scale.  This is not inconsistent with previous remarks, but given the speculative nature of our proposed $dS_4$ solution, it may be premature to try to make a definite connection with cosmology.

\section*{Acknowledgements}

The work of SSG~was supported in part by the Department of Energy under Grant No.\ DE-FG02-91ER40671, and by the Sloan Foundation.

IM wishes to thank O.~Ganor and M.~Rangamani for useful discussions. IM's 
work was supported in part by the Director, Office of Science,  
 Office of High Energy and Nuclear Physics, of the U.S. Department of  
 Energy under Contract DE-AC03-76SF00098, and in part by the NSF under  
 grant PHY-0098840 and grant PHY-0244900.

JF would like to thank J.~McGreevy and G.~Michalogiorgakis for insightful 
comments.  JF's work was funded in part by the NSF Graduate Research Fellowship 
Program.

\bibliographystyle{ssg}
\bibliography{dilaton}

\begingroup\raggedright\begin{thebibliography}{10}

\bibitem{clw}
E.~J. Copeland, A.~Lahiri, and D.~Wands, ``Low-energy effective string
  cosmology,'' {\em Phys. Rev.} {\bf D50} (1994) 4868--4880,
  \href{http://xxx.lanl.gov/abs/hep-th/9406216}{{\tt hep-th/9406216}}.

\bibitem{ekpyrosis1}
J.~Khoury, B.~A. Ovrut, P.~J. Steinhardt, and N.~Turok, ``The ekpyrotic
  universe: Colliding branes and the origin of the hot big bang,'' {\em Phys.
  Rev.} {\bf D64} (2001) 123522,
  \href{http://xxx.lanl.gov/abs/hep-th/0103239}{{\tt hep-th/0103239}}.

\bibitem{ekpyrosis2}
J.~Khoury, B.~A. Ovrut, N.~Seiberg, P.~J. Steinhardt, and N.~Turok, ``From big
  crunch to big bang,'' {\em Phys. Rev.} {\bf D65} (2002) 086007,
  \href{http://xxx.lanl.gov/abs/hep-th/0108187}{{\tt hep-th/0108187}}.

\bibitem{lmCreate}
A.~E. Lawrence and E.~J. Martinec, ``String field theory in curved spacetime
  and the resolution of spacelike singularities,'' {\em Class. Quant. Grav.}
  {\bf 13} (1996) 63--96, \href{http://xxx.lanl.gov/abs/hep-th/9509149}{{\tt
  hep-th/9509149}}.

\bibitem{gCreate}
S.~S. Gubser, ``String production at the level of effective field theory,''
  \href{http://xxx.lanl.gov/abs/hep-th/0305099}{{\tt hep-th/0305099}}.

\bibitem{clwReview}
J.~E. Lidsey, D.~Wands, and E.~J. Copeland, ``Superstring cosmology,'' {\em
  Phys. Rept.} {\bf 337} (2000) 343--492,
  \href{http://xxx.lanl.gov/abs/hep-th/9909061}{{\tt hep-th/9909061}}.

\bibitem{GR}
I.~Gradshteyn and I.~Ryzhik, ``Table of Integrals, Series, and Products,''.
  Academic Press, Inc. (1965) 1082p.

\bibitem{lebedev}
N.~Lebedev, ``Special Functions and Their Applications,''. Prentice-Hall, Inc.
  (1965) 308p.

\bibitem{gTalk}
S.~S. Gubser, ``String creation and cosmology,''
  \href{http://xxx.lanl.gov/abs/hep-th/0312321}{{\tt hep-th/0312321}}.

\bibitem{tsujikawa}
S.~Tsujikawa, ``Construction of nonsingular cosmological solutions in string
  theories,'' {\em Class. Quant. Grav.} {\bf 20} (2003) 1991--2014,
  \href{http://xxx.lanl.gov/abs/hep-th/0302181}{{\tt hep-th/0302181}}.

\bibitem{vt}
A.~A. Tseytlin and C.~Vafa, ``Elements of string cosmology,'' {\em Nucl. Phys.}
  {\bf B372} (1992) 443--466,
  \href{http://xxx.lanl.gov/abs/hep-th/9109048}{{\tt hep-th/9109048}}.

\bibitem{bv}
R.~H. Brandenberger and C.~Vafa, ``Superstrings in the early universe,'' {\em
  Nucl. Phys.} {\bf B316} (1989) 391.

\end{thebibliography}\endgroup

\end{document}